\documentclass{article}

\usepackage{english} 

\usepackage{psfig}
\usepackage{chicago}

\usepackage[dvips]{graphicx} 

\usepackage{proceedings}

\begin{document}

\def\rz{\ifmmode{I\hskip -3pt R}
    \else{\hbox{$I\hskip -3pt R$}}\fi} 

\title{A condition for the genotype--phenotype mapping: Causality}
\author{{\bf Bernhard Sendhoff}\thanks{~\{bs,kreutz\}@neuroinformatik.ruhr-uni-bochum.de}
\And Martin Kreutz\\
  Institut f\"ur Neuroinformatik, Ruhr-Universit\"at Bochum\\
  44780 Bochum, Germany\\
\And Werner von Seelen}

\maketitle

\vspace*{-9.5cm}\hfill\\
\hspace*{-1cm}
\begin{minipage}[t]{10cm}
{\bf published in:\\}
T. B{\"a}ck (Ed.) (1997)
{\em Proceedings of the Seventh International Conference 
on Genetic Algorithms \mbox{(ICGA'97)}}, Morgan Kauffman, 73-80.
\end{minipage}
\vspace*{6.5cm}\hfill\\

\begin{abstract}
The appropriate choice of the 
genotype $\rightarrow$ phenotype mapping in combination with the
mutation operator is important for a successful evolutionary search 
process. We suggest a measure to quantify the quality of this combination 
by addressing the question whether the relation among distances is
carried over from one space to the other. Search processes which do
not destroy the neighbourhood structure are termed {\em strongly causal}.
We apply the proposed measure to parameter and structure optimisation
problems in order to assess the combination (mapping, mutation
operator) and at the same time to be able to propose improved settings.
\end{abstract}

\section{Introduction}
\vspace*{-1ex}
\label{sec1}
The optimisation process in evolutionary algorithms is largely
influenced by the mapping from the genotype space to the phenotype
space. Especially for structure optimisation problems a measure of 
the quality of  the combination (mapping, mutation,
crossover) would be desirable. 
In this paper we propose such a measure based upon the observation
that Darwinian evolution takes gradual changes to the optimum, although 
in biological evolution other phenomena like punctuated equilibria
are also observed.

\begin{figure}[ht]
\centerline{
\includegraphics[width=\columnwidth]{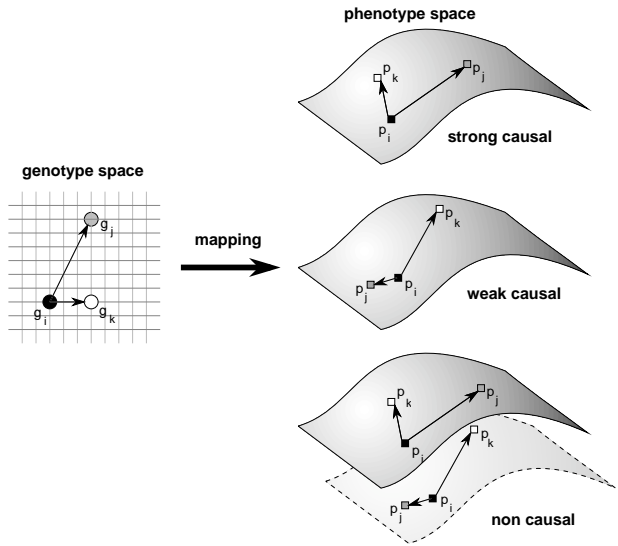}}
\caption[cau]{\label{causalFig}Examples for strongly, weakly and non
  causal genotype--phenotype mappings under the influence of mutation.
Circles denote genotypes $(g_i,g_j,g_k)$, $g_j$ and $g_k$ are results
of mutations from $g_i$. The corresponding phenotypes $(p_i,p_j,p_k)$ 
are shown as squares. The first strongly
causal mapping does not destroy the neighbourhood structure in
genotype space, the second weakly causal
mapping, maps small mutations in genotype space to large distances in
phenotype space and vice versa. The last example shows a non-causal
mapping.}
\end{figure}

We demand that the search process is {\em locally} strongly causal 
with respect to the mutation operator, that is: {\em small variations 
on the genotype space due to mutation imply small variations in the 
phenotype space}. This way the neighbourhood 
structure under the mapping ${\cal G} \rightarrow {\cal P}$ is
conserved, see Figure
\ref{causalFig}. The distance on the genotype
space is defined via the mutation probability. The need for a strong 
causal exploration of the search space has been expressed before
\shortcite{Rechenberg:94,Lohmann:93}. However, in the following we want to 
quantify the degree to which the setting (mapping, mutation operator)
satisfies the causality condition. 

The distance measure and therefore the causality condition in section
\ref{sec2} only depends on the mutation and not on the
crossover operator. This does not represent any opinion whether one or
the other is the driving force in evolutionary algorithms. However,
we believe that the mutation operator usually is responsible for
small steps in the phenotype space, hence for gradual changes
which we want to analyse. Furthermore, we assume a 
locally smooth fitness function and define conditions for the
genotype $\rightarrow$ phenotype mapping for this problem domain.
Thus, unlike in correlation based analysis,
\shortcite{Jones:95,Manderick:91}, we do not explicitly refer to a fitness
landscape, instead we focus on the conservation of neighbourhood
structures. 

In the next section we will propose a condition for a strongly causal
search process and quantify it by introducing a probabilistic
interpretation of the condition. Section \ref{sec3} presents
a first application in the domain of parameter optimisation problems and
the following section is concerned with the structure optimisation of
neural networks, where complicated genotype $\rightarrow$ phenotype
mappings are commonly used.

\section{A Condition for Causality}
\vspace*{-1ex}
\label{sec2}
\begin{minipage}{\columnwidth}
In section \ref{sec1} we claim that for the successful introduction of
new information by mutation the mutation operator should preserve the
neighbourhood structure in the corresponding evolutionary spaces.
We believe that strong causality is necessary 
in evolutionary algorithms 
\end{minipage}
\begin{itemize}
  \item to allow for controlled small steps in the phenotype space which
        are provoked by small steps in the genotype space.
        Especially in the vicinity of an optimum we need small steps to
        gradually approach the optimum.
  \item for the ability of self-adaptation of any strategy parameters, since
        with the lack of strong causality the information about the past
        is meaningless and adaptation is impossible.
\end{itemize}
In order to formulate the causality condition we have to define the term
{\em small variation\/} in a mathematical sense.
Therefore, we introduce a measure of distance in the genotype and phenotype
space.
For the mathematical correctness we have to show that the measure in the
respective spaces endows these spaces with a metric.
For distances in the genotype space we propose a ``universal'' measure
which is based on the probability of reaching genotype $g_j$ from genotype
$g_i$.
In this respect it resembles definitions of distance used in evolutionary
biology, \shortcite{Schuster:95b}:
{\em ``\ldots the notion of distance in genotype space is given by
the smallest number of individual mutations required for the
interconversion of two genotypes \ldots''}.
Furthermore, this measure is general enough to be applicable to a wide range
of evolutionary algorithms.\\
We introduce the following notations:
{\em Genotype space} ${\cal G}=\{g_i\}$ and 
{\em phenotype space} ${\cal P}=\{p_i\}$.
Both ${\cal G}$ and ${\cal P}$ can also be continuous spaces.
The mapping between the spaces is $f:{\cal G}\mapsto{\cal P}$, thus 
$p_i = f(g_i)$. The operators
mutation and crossover act upon the space ${\cal G}$, the
selection operator acts upon the fitness space ${\cal F}$
and therefore on ${\cal P}$.
We parameterise the mutation operator by a real valued vector 
$\vec{\sigma}\in \rz^l$.

Now, we will introduce the definition of distance on ${\cal G}$,
based on the mutation probability
$P(g_i\! \stackrel{\vec{\sigma}}{\rightarrow}\! g_j)$
of reaching $g_j$ from $g_i$ in ${\cal G}$
via mutation which is characterised by $\vec{\sigma}$.
\begin{eqnarray}
\label{e:geno-distance}
d(g_i,g_j) & = & - \log\left(\frac{1}{P_{id}}\,P(g_i\!
\stackrel{\vec{\sigma}}{\rightarrow}\! g_j)
\right) \\
\label{norm}
P_{id} & = & P(g\! \stackrel{\vec{\sigma}}{\rightarrow}\! g)
\end{eqnarray}
This definition is only sensible if we claim that
$P(g_i\!\stackrel{\vec{\sigma}}{\rightarrow}\! g_j) < P_{id}$ and that
the probability not to mutate is independent of $g$,
which is satisfied by most evolutionary
algorithms\footnote{
In GAs $P(g_i\!\stackrel{\vec{\sigma}}{\rightarrow}\! g_j) < P_{id}$
corresponds to a mutation rate $p<0.5$ ($p=0.5$ leads to random
initialisation) and in ES 
to normally distributed mutations with zero mean.}.
The logarithm in eq.\ (\ref{e:geno-distance}) is introduced in
order to make the distance measure additive instead of multiplicative.
The properties of this measure are discussed in \shortcite{Sendhoff:97a}.

Eq.\ (\ref{e:geno-distance}) allows for the comparison
between different EAs independent of any particular
metric on the genotype space, like Hamming distance or Euclidian distance. 

Now, we can proceed with the definition of causality.

\noindent{\bf Condition}: {\em Strong causality}\\[1ex]
\centerline{$\displaystyle
\forall g_i,g_j,g_k \;\; \exists \vec{\sigma'}, \varepsilon
~~\mbox{with}~~
\vec{\sigma} \in U_{\varepsilon}(\vec{\sigma'})$}
\begin{eqnarray}
\label{e:causal}
&&||f(g_i) - f(g_j)|| < ||f(g_i) - f(g_k)|| \nonumber\\
 & \Longleftrightarrow &   - \log\!\left(\frac{P(g_i\!\!
\stackrel{\vec{\sigma}}{\rightarrow}\!\! g_j)}{P_{id}}\right) < 
- \log\!\left(\frac{P(g_i\!\!
\stackrel{\vec{\sigma}}{\rightarrow}\!\! g_k)}{P_{id}}\right)\nonumber\\
 & \Longleftrightarrow & 
P(g_i\!
\stackrel{\vec{\sigma}}{\rightarrow}\! g_j) >  P(g_i\!
\stackrel{\vec{\sigma}}{\rightarrow}\! g_k)
\end{eqnarray}
The additional condition that $\vec{\sigma}$ can be drawn from
anywhere inside a sphere with radius $\varepsilon$ 
($\varepsilon$ can be sufficiently small) around $\vec{\sigma'}$ 
guarantees that the effect of mutation {\em continuously} 
varies with $\vec{\sigma}$.
That is, besides the existence of an appropriate $\vec{\sigma}$,
we have to guarantee that it is possible to locate.
Mathematically, 
the space of all mutation parameters which satisfy the causality
condition is not empty {\bf and additionally} not of measure zero.

We have indicated, that our analysis is concerned with
the local behaviour of evolutionary search.
Therefore, condition (\ref{e:causal}) should not be seen as a 
global condition. The term {\em local\/} is difficult to define.
However, an absolute measure of locality is not necessary since we are
interested in the relative performance of EAs.

Condition (\ref{e:causal}) defines strong causality in both
directions. Small distances and variations on the phenotype space
imply small distances and variations in the genotype space with respect
to the probability of jumping this distance via mutation and vice
versa.
However, in most EAs the second direction is more important.
That is, small variations in the genome provoke small variation in the
phenotype.

So far we have only set up a qualitative condition for strong causality.
In order to compare between EAs, we have to find a quantitative version of
condition (\ref{e:causal}).
We will rephrase it in the light of a probabilistic
interpretation.

Assuming the $g_i,g_j,g_k$ to be random variables with uniform distribution,
both sides of condition (\ref{e:causal}) become boolean random variables.
As a shortcut, we introduce the symbols $A$ and $B$:
\begin{eqnarray}
A &:=& ||f(g_i) - f(g_j)|| < ||f(g_i) - f(g_k)|| \\
\label{BEquals}
B &:=& - \log\!\left(\frac{P(g_i\!\!
\stackrel{\vec{\sigma}}{\rightarrow}\!\! g_j)}{P_{id}}\right) <
- \log\!\left(\frac{P(g_i\!\!
\stackrel{\vec{\sigma}}{\rightarrow}\!\! g_k)}{P_{id}}\right)\nonumber\\
& \Longleftrightarrow& B\; := \; 
P(g_i\! \stackrel{\vec{\sigma}}{\rightarrow}\! g_j) >
P(g_i\! \stackrel{\vec{\sigma}}{\rightarrow}\! g_k)
\end{eqnarray}
Since we assume the distribution of $g_i,g_j,g_k$ to be known, we can derive
the probabilities $P(A)$, $P(B)$, and $P(A,B)$.
We can now, with the help of Bayes' law, recast the two directions (genotype
$\leftrightarrow$ phenotype) in the following way:

\medskip

{\bf Probabilistic condition}: {\em Strong causality}\\[1ex]
\centerline{$\displaystyle
\forall g_i,g_j,g_k \;\; \exists \vec{\sigma'}, \vec{\varepsilon}
~~\mbox{with}~~
\vec{\sigma} \in U_{\varepsilon}(\vec{\sigma'})$}
\begin{eqnarray}
{\cal G} \Rightarrow {\cal P}:~~
\label{PAB}
P(A|B) &=& \frac{P(A, B)}{P(B)}\, = \, 1\\
\label{PBA}
{\cal P} \Rightarrow {\cal G}:~~
P(B|A) &=& \frac{P(A, B)}{P(A)}\, = \, 1
\end{eqnarray}
The value of $P(A|B)$ serves 
as a quantitative measure for the causality in EAs.
If the neighbourhood relations in both spaces are uncorrelated for
every point, then the system is weakly but {\bf not} strongly causal 
$P(A,B) = P(A)\cdot P(B)$ and therefore $P(A|B)=P(A)$, $P(B|A)=P(B)$,
thus distance relations 
in the phenotype space are statistically independent from distance
relations in the genotype space and vice versa\footnote{Whether
the system is non-causal in the sense of being non-deterministic
is not determined by 
eqs.\ (\ref{e:causal},\ref{PAB},\ref{PBA}), since we do not
observe whether the mapping from genotype to phenotype space itself
is probabilistic or not.}.
One example for such systems is the class of Monte Carlo algorithms
where the transition probability between any pair of genotypes is
constant.
For constant 
transition probabilities, $B$ in equation (\ref{BEquals}) is constant for
all genotype combinations and does therefore not provide any information 
about the distance relation in the phenotype space.

In evolutionary molecular biology
measures similar to this
probabilistic formulation of the causality condition are employed in the
context of the analysis of the ``sequence--structure'' mapping
\shortcite{Schuster:95a}.

\section{Parameter Optimisation}
\vspace*{-1ex}
\label{sec3}
Firstly, we employ one of the mainstream paradigms of EAs -- the evolution
strategy (ES) and show that the ES is strongly causal in terms of
our proposed condition.
As an example for an EA, which violates the causality condition,
we analyse the canonical genetic algorithm (GA) applied to parameter
optimisation.
We propose a new mutation operator for the GA which observes strong causality
to a greater extent and show that this also increases the
performance.

\subsection{Evolution Strategy}

We firstly focus on the transition probability.
In the canonical ES ${\cal G}\!=\!{\cal P}\!=\!\rz^n$ and the
genotype $\rightarrow$ phenotype mapping is the identity
\mbox{$f:{\cal G}\rightarrow {\cal P}$} $= {\mbox{id}}_{\rz^n}$.
It uses normally distributed mutation steps which
are independent of the genotype $g_i \in {\cal G}$.
That is, the transition $g_i\!\stackrel{\vec{\sigma}}{\rightarrow}\!g_k$ is
defined by adding a normally distributed number $z=(z_1,\dots,z_n)$ with
$z_i\sim N(0,\sigma^2)$.
Hence, the pdf of this transition can be expressed in terms of $z$
\begin{eqnarray}
    g_i\!\!\stackrel{\vec{\sigma}}{\rightarrow}\!\!g_k & : & g_j = g_i + z \\
    p(g_i\!\!\stackrel{\vec{\sigma}}{\rightarrow}\!\!g_k) & = & p(z = g_j - g_i )
    \nonumber\\
    & = & \frac{1}{\sqrt{2\pi}^{\,n}\sigma^n}
       \exp\!\left(\!-\frac{\|g_k\!\!-\!\!g_i\|^2}{2 \sigma^2}\!\right) \\
    p_{id} & = & p(z=0) ~=~ \frac{1}{\sqrt{2\pi}^{\,n}\sigma^n}
\end{eqnarray}
Inserting the transition pdf in the causality condition (\ref{e:causal})
with $f:{\cal G}\rightarrow {\cal P} = {\mbox{id}}_{\rz^n}$ results in
\begin{eqnarray}
& & \|f(g_i)-f(g_j)\| < \|f(g_i)-f(g_k)\|\nonumber\\
& \Longleftrightarrow &
\|g_i-g_j\| < \|g_i-g_k\| \nonumber\\
& \Longleftrightarrow &
\exp\left(-\|g_i-g_j\|^2\right) > \exp\left(-\|g_i-g_k\|^2
\right)\nonumber\\
& \Longleftrightarrow &
p(g_i\!\!
\stackrel{\vec{\sigma}}{\rightarrow}\!\!g_j) >
p(g_i\!\!
\stackrel{\vec{\sigma}}{\rightarrow}\!\!g_k)
\end{eqnarray}
which holds for all combinations of $g_i,g_j,g_k,\sigma$.

The examination of the metric conditions of the distance measure of an
ES and some notes on the self-adaptation of $\sigma$
are presented in \shortcite{Sendhoff:97a}.

\subsection{Genetic Algorithms}

In the case of genetic algorithms (GA) the genotype space consists
of binary strings of length $L$, therefore ${\cal G}=\{0,1\}^L$.
Canonical GAs mutate by changing each bit position from
$0\rightarrow 1$ and $1\rightarrow 0$, respectively,
with the probability $p_m$.
Thus, $p_m$ corresponds to the mutation parameter $\sigma$.
Let $h_{ij}$ denote the Hamming distance between $g_i$ and $g_j$.

In order to examine the causality condition we
use the Euclidian metric on the phenotype space ${\cal P}$ and
choose the standard binary coding for the genotype--phenotype mapping
$f:{\cal G}\rightarrow{\cal P}$.
Using the following notations
\begin{eqnarray}
g_i & = &\{x_i(n)\,|\,x_i(n) \in \{0,1\} \}\\
h_{ij} & = & \sum_{n=0}^{L-1} \left|x_i(n)-x_j(n)\right| \\
f(g_i) & = & \sum_{i=0}^{L-1} x_i(n) \,2^n \\
P(g_i\!\stackrel{\vec{\sigma}}{\rightarrow}\!g_j) & = &
p_m^{h_{ij}} (1-p_m)^{L-h_{ij}}
\end{eqnarray}
the causality condition is expressed as
\begin{eqnarray}
& \displaystyle\left| \sum_{n=0}^{L-1} (x_i(n)\!-\!x_j(n)) \,2^n \right| <
\left| \sum_{n=0}^{L-1} (x_i(n)\!-\!x_k(n)) \,2^n \right| & \nonumber\\[1ex]
& \displaystyle\Longleftrightarrow~
p_m^{h_{ij}} (1\!-\!p_m)^{L\!-\!h_{ij}} > p_m^{h_{ik}} (1\!-\!p_m)^{L\!-\!h_{ik}}~~~ &
\label{gaCaus1}
\end{eqnarray}
Assuming $p_m<0.5$ the right hand side of (\ref{gaCaus1}) can be
expressed as $h_{ij}<h_{ik}$. Therefore the causality condition reads
\begin{eqnarray}
& \displaystyle\left| \sum_{n=0}^{L-1} (x_i(n)\!-\!x_j(n)) \,2^n \right| <
\left| \sum_{n=0}^{L-1} (x_i(n)\!-\!x_k(n)) \,2^n \right| & \nonumber\\
& \displaystyle\Longleftrightarrow~
\sum_{n=0}^{L-1} \left|x_i(n)\!-\!x_j(n)\right| <
\sum_{n=0}^{L-1} \left|x_i(n)\!-\!x_k(n)\right|~~~~~~ &
\label{Cond1GA}
\end{eqnarray}
which obviously does not hold in general, not even locally.

As a measure of the extent to which the GA
satisfies the causality condition we employ the probabilistic version of
the condition.
After some extensive calculations which are presented in \shortcite{Sendhoff:97a}
we get \mbox{$P(A|B)\approx 0.51$} and \mbox{$P(B|A)\approx 0.62$}.
That is, the chance of a small mutation of a genotype resulting
in a small change of the corresponding phenotype is about 51\%.
The probability that a small change of a phenotype is caused by a small
mutation of the genotype is somewhat higher, about 62\%.
Thus, in the case of a canonical GA, the mapping from the genotype
to the phenotype is not strongly causal.
In our opinion the combination of binary coding and point mutation
is not well suited for continuous parameter optimisation combined
with locally smooth fitness functions and is the reason
why ES, which observes strong causality, outperforms the GA in 
most cases in this problem domain.

\subsection{A New Mutation Operator}

We have seen that in GA the standard mutation operator together with the
binary encoding does not satisfy the causality condition in general.
Possible solutions to this problem are to use a different encoding scheme,
e.g.\ the Gray code\footnote{Although we show in \shortcite{Sendhoff:97a}
that the Gray code does not increase the causality substantially.},
to change the mutation operator and
keep the encoding scheme, and to change both.

In the remainder of this section we will partly outline an approach,
presented in detail 
in \shortcite{Sendhoff:96}, which sticks to the concept of point mutation,
but uses a position dependent mutation rate $p_m=p_m(i)$.
This will provide us with
an interesting example of an EA, where a modification of the mutation
operator enhances the causality and,
as we will see, also the performance.

\begin{figure}[htb]
\centerline{\psfig{file=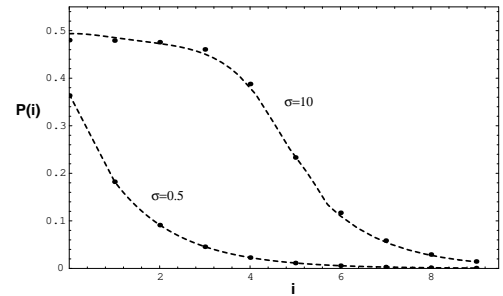,width=0.78\columnwidth}}\vspace*{-1ex}
\caption[]{\label{f:pmi}
Top: $p_m(i)$ for $\sigma=10.0$ 
(dashed curve - numerical approximation),
bottom: $p_m(i)$ (rescaled with a factor 8) 
for $\sigma=0.5$ 
(dashed curve - numerical approximation).}
\end{figure}

As we have seen above, the ES is a strongly causal optimisation procedure.
Therefore, we translate the concept of mutation by adding normally
distributed numbers in ES to point mutation in GAs.
Thus, we calculate a probability distribution which will on average
resemble the summation of a normally distributed number.
Depending on the standard deviation $\sigma$ of the underlying normal
distribution we get different distributions of the mutation rates $p_m(i)$,
see Figure \ref{f:pmi}.
For the efficient use of the new mutation operator we derived a numerical
approximation of $p_m(i;\sigma)$ which is presented in \shortcite{Sendhoff:96}.

\begin{figure}[htb]
\centerline{
(a) \psfig{file=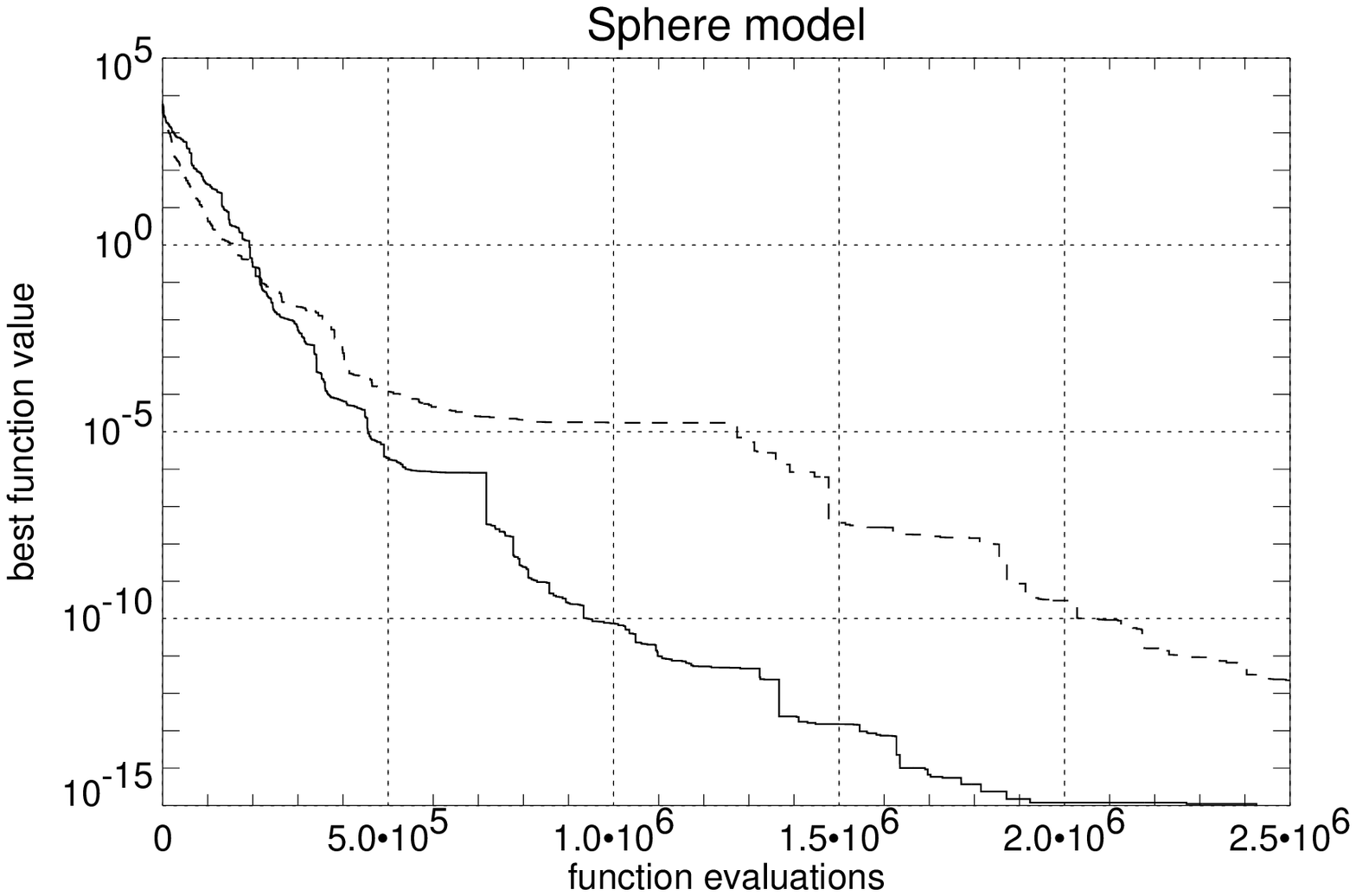,width=0.8\columnwidth}}
\centerline{
(b) \psfig{file=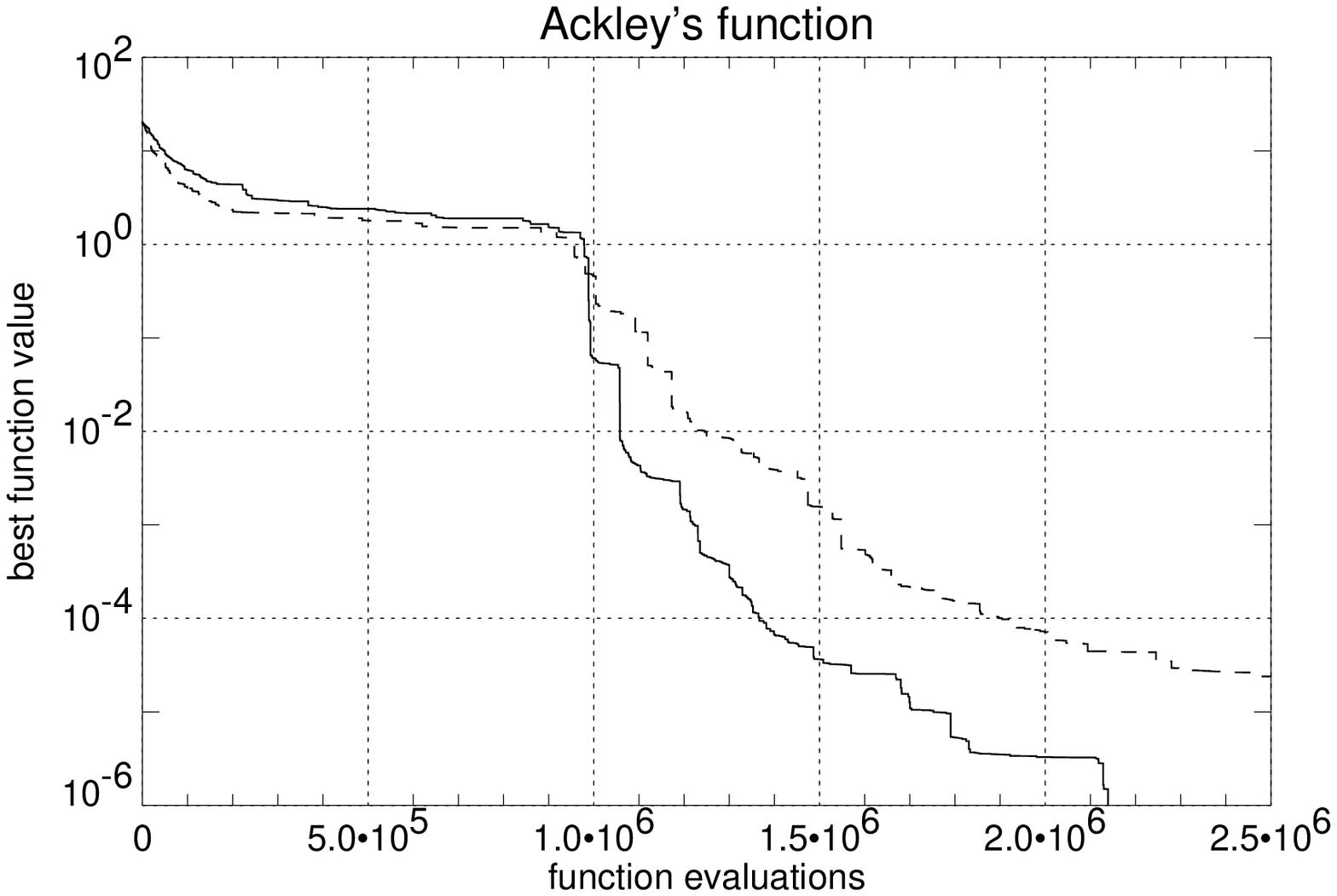,width=0.8\columnwidth}}
\caption[]{
\label{f:opt-results}
Convergence plots of optimisation runs. The solid line shows the results
obtained by position-dependent mutations. Dashed lines the ones from
the canonical GA. 
(population size: 50; dimension: $n=30$; encoding length: 32 {\tt bits} using Gray-code)}
\end{figure}

The numerical estimates of the causality measure are 
$P(A|B)\approx 0.73$ and $P(B|A)\approx 0.74$.
In order to support our hypothesis that increasing the strong causality in
an optimisation process leads to an improved performance
we apply the modified GA to two standard optimisation
problems.
Results are given for the sphere model
and Ackley's function, see
figure \ref{f:opt-results}.
The new GA converges faster to a better value than the canonical GA.
Figure \ref{f:opt-results} shows that in case of the sphere model the increase
of convergence speed is of order $10^5$ and for the Ackley function of
order $10^{1.5}$.

\section{Causality in Structure Optimisation}
\vspace*{-1ex}
\label{sec4}
The problem to choose the right genotype $\rightarrow$ phenotype mapping  
is of particular importance in the domain of structure
optimisation. We will here concentrate on the
structure optimisation of neural networks.
We will regard the set of all possible connection matrices as the
phenotype space, and allow the matrix to have entries from $\{0,1\}$
or from $\{0,...,N_{sym}\}$. Since there is no measure on the space of 
these matrices which relates directly to the performance of the neural
network without evaluating the network, we will use the standard
Euclidian distance measure for the phenotype space, ($y_{nk}^{M_j}$
denotes the entry at (row $n$, column $k$) of the matrix $M_j$)
\begin{equation}
\label{dE}
d_E(M_i, M_j) = \frac{1}{N^2} \sum_{n=1}^{N} \sum_{k=1}^{N} |y_{nk}^{M_i}-
y_{nk}^{M_j}|
\end{equation}
There is no a priori structure assumed for the network, hence the matrix
is not constrained to any layered network structure. If $y_{nk} > 0$ 
there exists a connection between neuron $n$ and neuron $k$. We are
not restricted to the upper triangular part of the matrix, thus in
principle feedback connections can be specified. If the $y_{nk}$
are restricted to the values $\{0,1\}$, we only specify the
connection between the neurons. If we extend the allowed values
to all integers in the set $\{0,...,N_{sym}\}$, it is possible
to further define initial values for the weights and the thresholds.
In connection with gradient descent algorithms for the fine tuning
of the weights, this approach has been successful, see
\shortcite{Sendhoff:97}, and we will therefore include it in the following 
examinations.

There have been several proposals on how to organise the genotype space and
the mapping $f:{\cal G}\rightarrow{\cal P}$ for the optimisation
of the structure of neural networks, see also \shortcite{Whitley:95}. 
Most of them can be categorised
into three principal approaches, the {\em direct encoding}, the
{\em recursive or grammar encoding} and the {\em cellular encoding}. The
first attempts to use evolutionary (genetic) algorithms for
structure optimisation employed the {\em direct encoding} method, 
\shortcite{Miller:89}, and it probably still is the most frequently used method. 
The {\em recursive encoding} has been introduced by Kitano
\citeyear{Kitano:90},
in order to overcome the bad scaling behaviour of the 
{\em direct encoding} method for large networks and to favour a
modular structure of the network. The third
approach, the {\em cellular encoding}, proposed by Gruau \citeyear{Gruau:93}, 
uses a tree representation of operators which construct the network. 
The structure of the tree and therefore of the network is 
optimised by genetic programming. In the following we will examine 
the {\em direct
encoding} and the {\em recursive encoding} with respect to the
proposed measure, eqs.\ (\ref{PAB}, \ref{PBA}). 
Therefore, we will examine whether the neighbourhood
structure on ${\cal G}$ is carried over to ${\cal P}$; whether the 
system is strongly causal. We will restrict ourselves to the direction,
${\cal G}\rightarrow{\cal P}$ and we will not sample uniformly
in ${\cal G}$. The reason is, that for the 
mutation operator $p_{\pm}$ (see eq.\ (\ref{ppm})) 
the probability to reach the genotype $g_j$ and
$g_k$ from $g_i$ is zero for almost all uniformly sampled triples.
Thus, if we want to examine the system (mapping, $p_{\pm}$), we
sample $g_i$ uniformly and obtain $g_j$ and
$g_k$ via mutation from $g_i$ with the probability $p_{init}$.
By tuning $p_{init}$, we are at the same time able to determine how
{\em local\/} the three chromosomes are. We then derive the probability
to reach $g_j$ and $g_k$ from $g_i$ by ``normal mutation''. 

\subsection{The {\em direct encoding} method}
In the {\em direct encoding\/} method the chromosome consists of
the whole connection matrix. Usually all matrix rows are
concatenated to form the chromosome, whose elements we want to 
denote by $x_n$. The range of
allowed values for $x$ and $y$ can be $\{0,1\}$ or from the
integer set $\{0,...,N_{sym}\}$. The following 
operators ($\chi \in [0,1[$ is a uniformly sampled number) 
have been used
\begin{eqnarray}
\label{ppm}
p_{\pm} x  & = & \left\{\begin{array}{lrl}x+1&;&\chi<0.5\\x-1&;&\chi\geq 0.5
\end{array}\right.\\
p_{u} x & = & \left\lfloor \chi \cdot (N_{sym} +1) \right\rfloor
\end{eqnarray}
$p_{u}$ replaces $x$ by a new integer with equal probability 
from the set $\{0,...,N_{sym}\}$.
We used the Euclidian measure of distance for matrices 
on the phenotype space and a distance measure which only 
counts structural differences $d_I(M_i, M_j)$
\begin{eqnarray}
\label{dI}
&\displaystyle d_I(M_i, M_j)  =  \frac{1}{N^2} \sum_{n=1}^{N} \sum_{k=1}^{N}
|\Theta(y_{nk}^{M_i})-\Theta(y_{nk}^{M_j})|~~~~&\\
&\Theta(x)= \left\{\begin{array}{lrl}0&;&x\leq0\\1&;&x>0\end{array}\right.
&\nonumber
\end{eqnarray}
The results for the probability $P(A|B)$, that is the
probability that ($d_{E,I}$ denotes $d_{E}$ or $d_{I}$)
\begin{equation}
\label{B}
A:=d_{E,I}(M_i, M_j) > d_{E,I}(M_i, M_k)
\end{equation}
holds in phenotype space, given that 
\begin{equation}
\label{A}
B:=-\log\left(P(g_i\rightarrow g_j)\right) > 
-\log\left(P(g_i\rightarrow g_k)\right)
\end{equation}
is true in genotype space, are presented in Table \ref{direct}.
\setlength\tabcolsep{0.175cm}
\begin{table}[htb]
  \begin{center}
    \begin{tabular}{|l||c|c|c|c|}
      \hline
      &  $d_E$/$p_{\pm}$ & $d_I$/$p_{\pm}$ &
      $d_E$/$p_{u}$ & $d_I$/$p_{u}$ \\
      \hline\hline 
      $x \in \{0,1\}$ & -- & -- & -- & 0.0 \\
      \hline
      $x \in \{0,..,N_{sym}\}$ & 0.0 & 0.614 &  0.662 & 0.564 \\
      \hline
    \end{tabular}
  \end{center}\vspace*{-1ex}
\caption[x]{\label{direct}
  Numerical estimation of the probabilities $P(A|B)$, using
  combinations of the two different distance measures and mutation
  operators. The probabilities have been estimated from $10^5$
  trials ($N_{sym}=10$ and $p_{init}=0.25$).}
\end{table}
The standard setting, $(x \in \{0,1\}, d_I, p_{u})$ is strongly causal
in the ${\cal G}\rightarrow{\cal P}$ direction. However, if the 
allowed values are extended to an interval of integers, all
settings have problems at least for the structural distance measure.
Thus, we conclude that even {\em direct encoding methods} are not
strongly causal straightforwardly if we depart from the basic setting.

\subsection{The {\em recursive encoding} method}
\begin{figure}
\centerline{
\includegraphics[width=0.75\columnwidth]{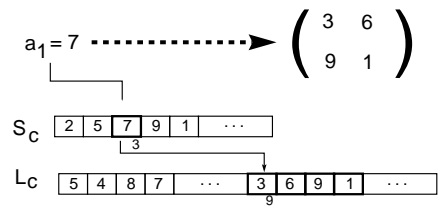}}\vspace*{-4ex}
\caption[]{\label{kitano1}One element is replaced by four elements in the
recursion step via the small chromosome $S_C$ $\rightarrow$ large chromosome 
$L_C$ mapping.}
\end{figure}

\begin{figure}
\centerline{
\includegraphics[width=\columnwidth]{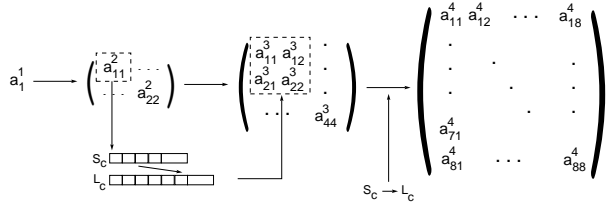}}\vspace*{-2ex}
\caption[]{\label{kitano2}Scheme of the recursive development of the
connection matrix up to a size of $8 \times 8$. 
}
\end{figure}

In all encoding methods apart from the one discussed above a 
more or less intrinsic mapping is introduced from the genotype
to the phenotype space. We already argued why this is
sensible and we now want to examine to what extent a 
{\em recursive encoding} method is strongly causal. The coding, 
described in \shortcite{Sendhoff:97}, consists of four chromosomes, where
only the first two are important for the building process of the
connection matrix. In each iteration step every element
of the connection matrix is replaced by a $2 \times 2$ matrix 
of new elements. 
The new elements are specified by a mapping from
the small chromosome $S_C$ to the large chromosome 
$L_C$. The length 
of the small chromosome $N_{S_C}$ is variable, the length of the 
large one is fixed by the condition $N_{L_C}=4\cdot N_{S_C}$. 
At each step $i$ the {\em first\/} place  $N(y_{nk}^i)$ of each 
connection matrix element $y_{nk}^i$ in $S_C$ is determined;
for example position $N(y_1=7)=3$ in Figure \ref{kitano1}.
The element is then replaced by the four elements at the positions 
\begin{eqnarray}
&\displaystyle
\big(\;4\cdot(N(y_{nk}^i)-1)+1,\; 4\cdot(N(y_{nk}^i)-1)+2,~~~~~~&\nonumber\\
\label{rec}
&\displaystyle
4\cdot(N(y_{nk}^i)-1)+3,\; 4\cdot(N(y_{nk}^i)-1)+4\;\big)~~&
\end{eqnarray}
in the large chromosome $L_C$.
Figure \ref{kitano1} shows the replacement of an element $y_1=7$
by the four elements $\left(3,6,9,1\right)$.
In case $y_{nk}^i$ is not in $S_C$, it is replaced by four so
called terminal symbols (in the notation of integer strings, the most
convenient choice is zero). A terminal symbol is in turn always replaced
by another four terminal symbols in an recursion step. 
Figure \ref{kitano2} shows the evolution of a $8\times 8$ connection matrix 
$M_{con}$ following the introduced rules.
This network connection matrix is a
function of the mutation and crossover probabilities, the 
chromosome length $d_{S_C}$, the number of iteration steps $N_{steps}$ and 
of the size of the set of integers $\{1,...,N_{sym}\}$ of allowed 
values for both strings. 
\begin{figure}[ht]
\centerline{
(a)
\includegraphics[width=0.8\columnwidth]{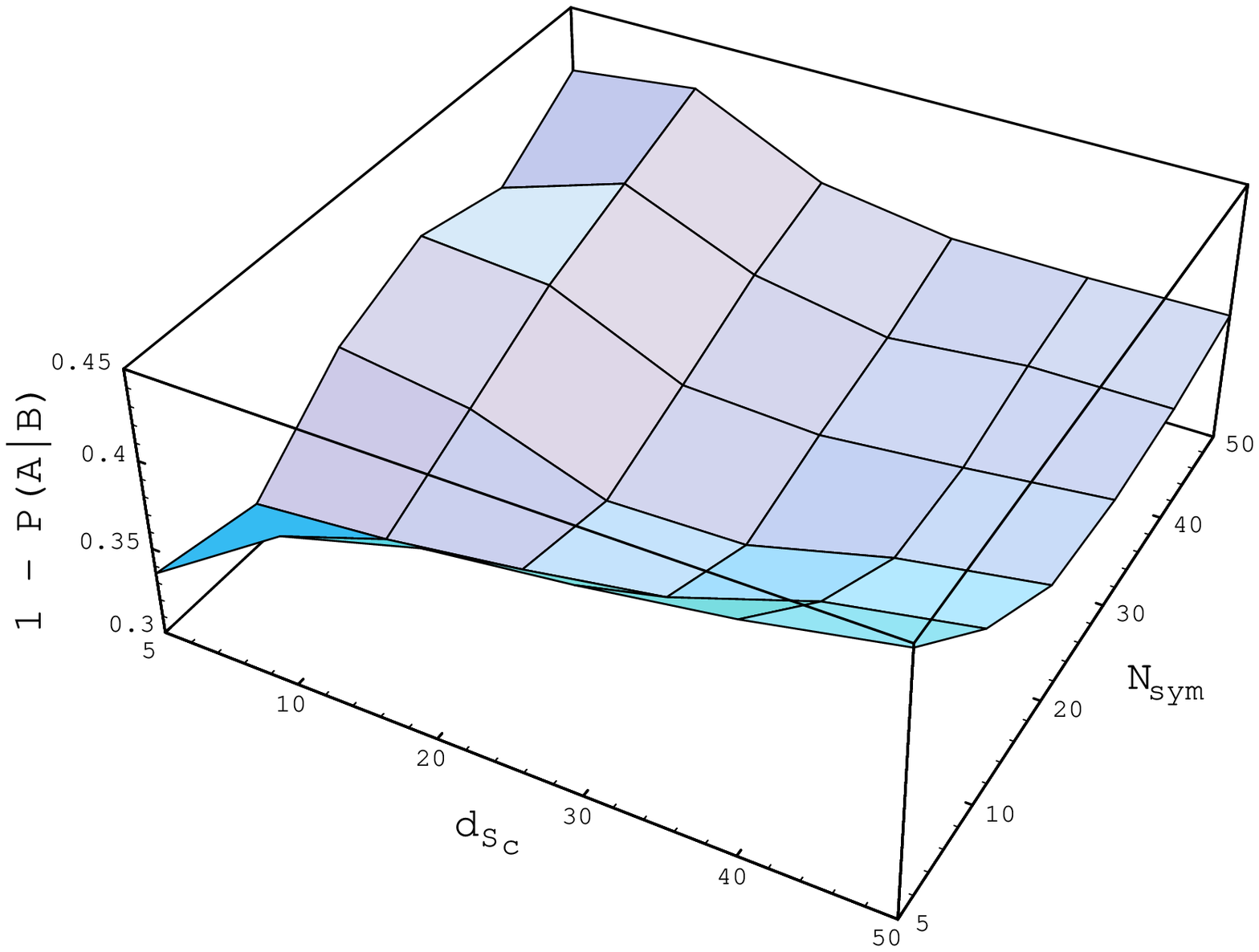}}\hfill\\
\centerline{
(b)
\includegraphics[width=0.8\columnwidth]{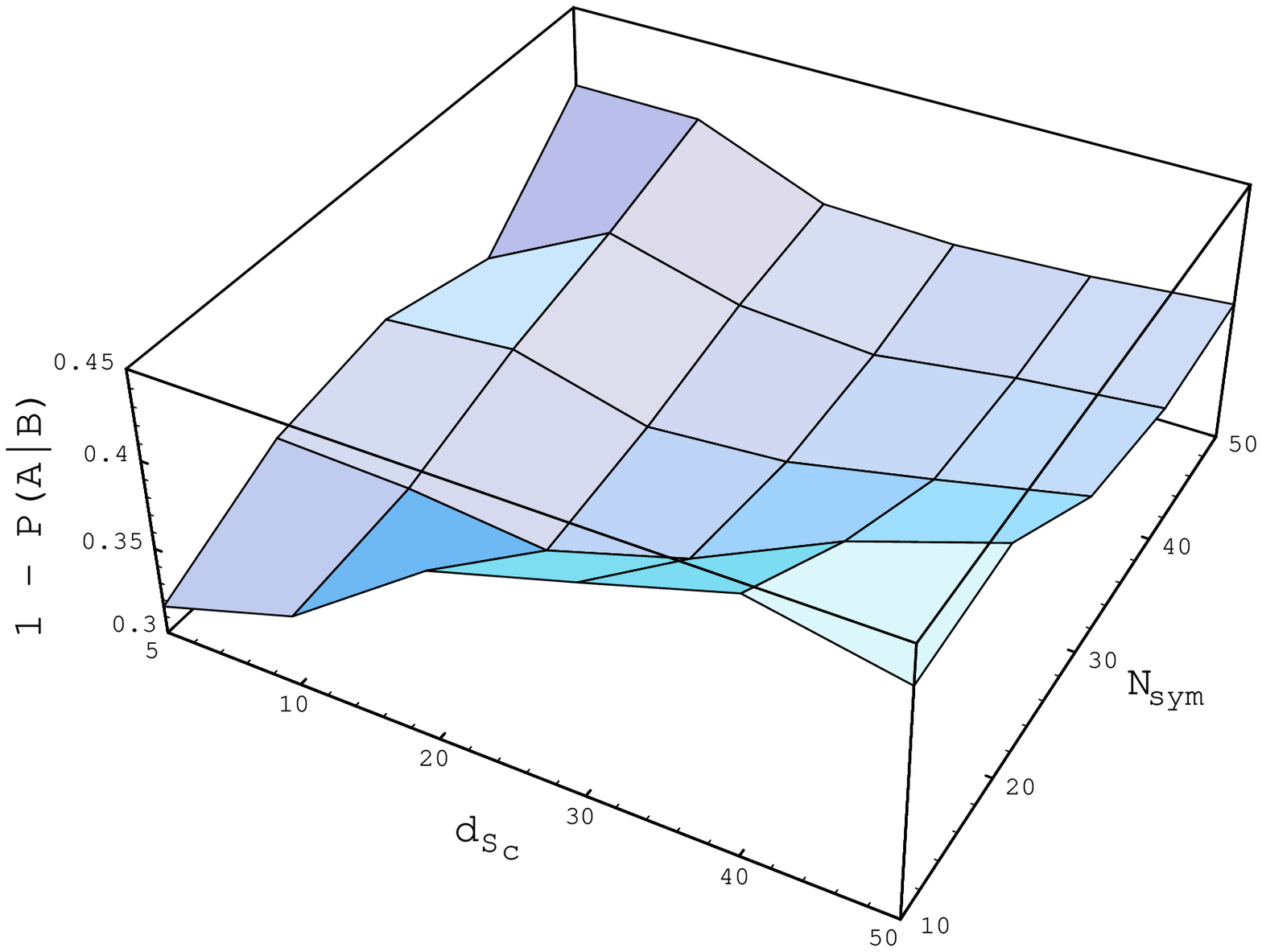}}
\caption[cau]{\label{Rec3D}Since it is easier to visualise, the 
  probability to {\bf violate} the causality
  condition $(1-P(A|B))$ is shown for the (a) Euclidian distance measure $d_E$
  and (b) structure distance measure $d_I$. The
  values have been estimated from $10^5$ trials ($p_{init}=0.25$).}
\end{figure}
We restrict ourselves to mutations on $S_C$
and examine the probability $(1-P(A|B))$ as
a function of $d_{S_C}$ and $N_{sym}$, the results are shown in 
Figure \ref{Rec3D} (a) for the Euclidian distance measure 
$d_E(M_i, M_j)$ and in (b) for the structure distance
measure $d_I(M_i, M_j)$. Only the results for the
$p_{\pm}$ mutation operator are shown, because the values for $p_u$
are only slightly lower and show a similar behaviour as in Figure
\ref{Rec3D}. We note that the system is generally
not strongly causal, especially for specific combinations ($d_{S_C},
N_{sym}$). Furthermore, the best (lowest, since causality {\bf
  violations} are shown) values on average are reached if
the encoding parameters $d_{S_C}$ and $N_{sym}$ differ only slightly,
thus we conclude $d_{S_C} \approx N_{sym}$. We also note that the
differences between $d_E$ and $d_I$ are only marginal both 
qualitatively and quantitatively. 
Thus, from the point of view of
causality the combined optimisation of the structure and the initial
weight values seems to be sensible.

\begin{figure}
\centerline{
(a)
\includegraphics[width=0.8\columnwidth]{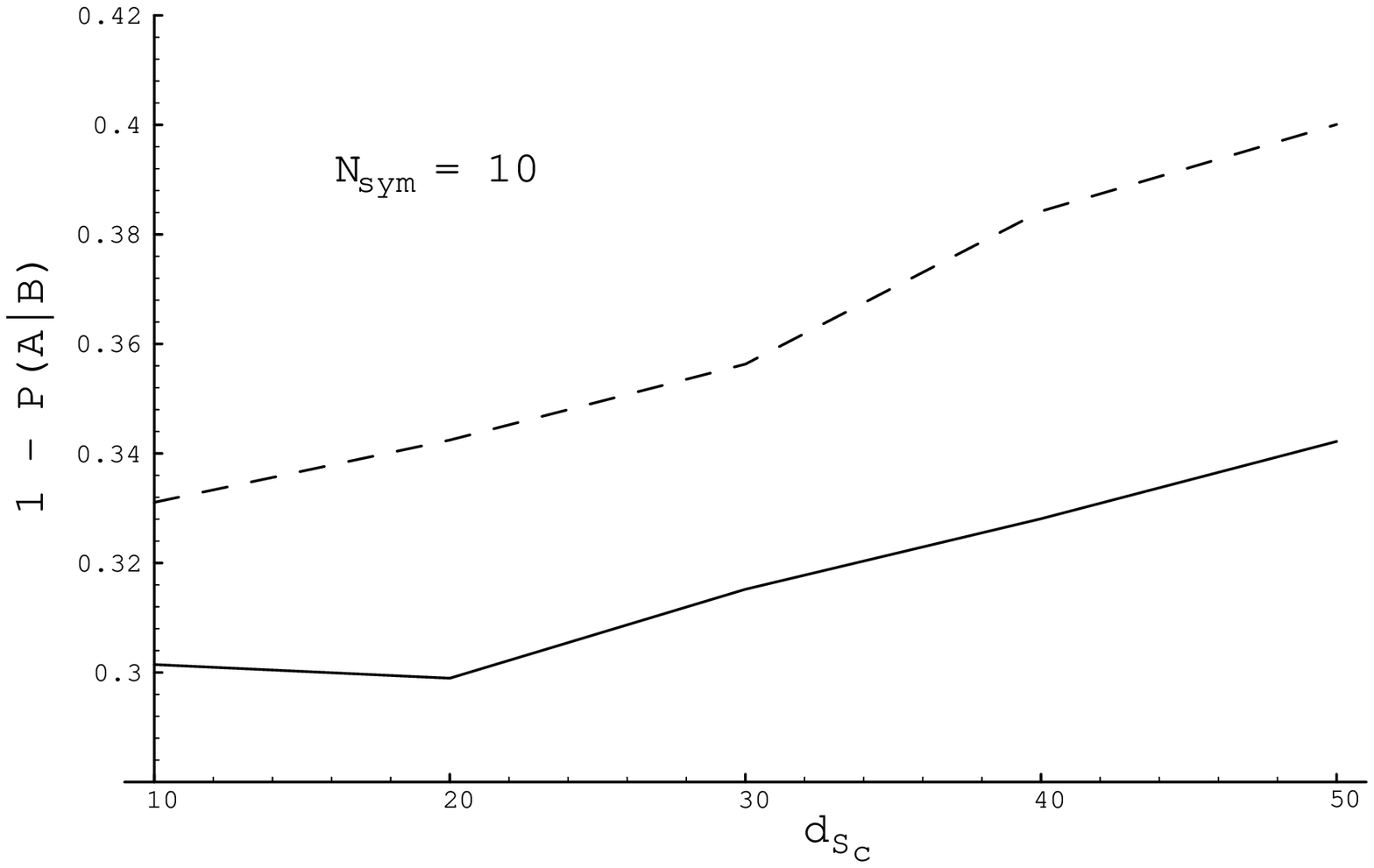}}\hfill\\
\centerline{
(b)
\includegraphics[width=0.8\columnwidth]{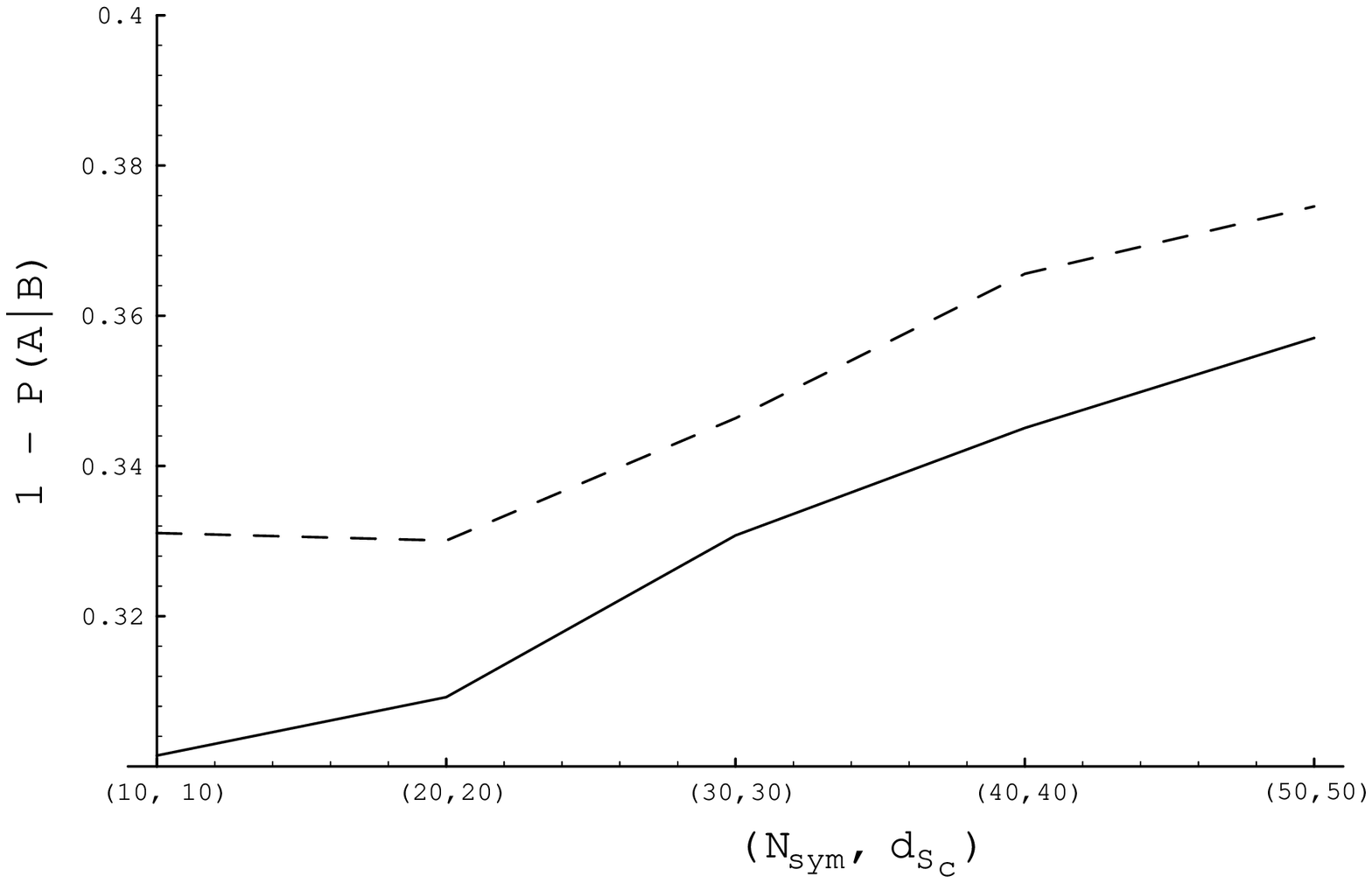}}
\caption[cau]{\label{Rec2D}The probability to {\bf violate} the causality
  condition $(1-P(A|B))$ (see also Fig.\ \ref{Rec3D}) 
  estimated from $10^5$ trials. (a) $N_{sym}=10$
  is kept constant and (b) the relation $N_{sym}=d_{S_C}$ is fixed.
  The interlaced curve shows the values for 
  the mutation operator with the modifications (\ref{mod1},
  \ref{mod2}) and the dashed curve for the standard mutation rate
  $p_m$ ($p_{init}=0.25$).}
\end{figure}

We will now, similar to section \ref{sec3} in the domain of parameter
optimisation, try to lower the probability of causality violations
with the help of a new position dependent mutation operator.
Firstly, we have to identify typical settings which are responsible for
causality violations. One is a direct consequence of the redundant
nature which from the viewpoint of accumulated mutation is also 
advantageous. Hence, we do not try to change the encoding to be less
redundant, but instead we change the mutation operator, so that the
mutation probability rises for redundant chromosome entries. If $p_m$
is the probability to mutate and $N_{x_k(S_C)}$ denotes the number of
occurrences of symbol $x_k(S_C)$ in $S_C$ before the position of $x_k(S_C)$,
we write 
\begin{equation}
\label{mod1}
p_m(x_k(S_C)) = p_m \cdot (N_{x_k(S_C)} + 1)
\end{equation}
Secondly, we observe that all elements from $S_C$ which occur in the 
first four elements in $L_C$ have a large impact on the connection
matrix, since the first element in $S_C$ is always mapped onto this first
block of elements in $L_C$. Therefore, we suggest a
second modification to the mutation operator:
\begin{eqnarray}
\label{mod2}
p_m(x_k(S_C)) & = & p_m^2 \nonumber\\
\forall \; x_k(S_C) & \in & \{x_1(L_C),\ldots,x_4(L_C)\}
\end{eqnarray}
Figure \ref{Rec2D} (a) and (b) show the results for the probability
of causality violating steps for the mutation operator with the
modifications eqs.\ (\ref{mod1}, \ref{mod2}) compared to
the fixed mutation (dashed curve) rate $p_m$. In Figure \ref{Rec2D} (a)
we kept $N_{sym}$ constant and changed the length of $S_C$, since we
expect that in this case modification (\ref{mod1}) will have the
largest impact because the amount of redundant elements in $S_C$
rises with $d_{S_C}$. Indeed, we observe that $(1-P(A|B))$ is
considerably reduced and that the effect is more pronounced
for larger values of $d_{S_C}$. 
Figure \ref{Rec2D} (b) shows
experiments carried out for the combinations $N_{sym}=d_{S_C}$ which,
as we pointed out earlier, are the best choices for the coding
parameters. The new mutation operator reduces the probability of
causality violations also in these cases, however the difference
to the fixed mutation rate is smaller than in  Figure \ref{Rec2D} (a).
Thus, we conclude that minor modifications of the mutation operator
can already have an causality enhancing effect on the search process
and that it is worthwhile to analyse the (genotype $\rightarrow$
phenotype, mutation) system with respect to the question why and for
which specific settings problems can occur.

\section{Conclusion}
\vspace*{-1ex}
\label{sec5}
In this paper we suggested a condition which the setting (genotype
$\rightarrow$ phenotype mapping, mutation) should fulfill in order to allow
gradual changes for a local search on the phenotype space 
which can be controlled via the mutation parameter on the genotype
space. 
We applied the probabilistic causality condition
to problems in the domain of parameter optimisation 
and structure optimisation both analytically and constructively. 
Thus, besides
examining the search process, we also 
suggested variations in the mutation operator to improve the
setting with respect to our condition. Especially in the later domain,
where complicated mappings are commonly used, we believe the measure
could be a useful tool for constructing evolutionary
algorithms. In the case of parameter optimisation, Figures 
\ref{f:opt-results} show that the setting which enhances the 
causal behaviour at the same time improves the performance. 
\vspace*{-1.3ex}
\subsubsection*{Acknowledgements} 
\vspace*{-1.5ex}
This research work is part of the BMBF {\em SONN}
project under Grant No. 01IB401A9.
\vspace*{-1ex}

{\renewcommand{\baselinestretch}{0.9995}
\small 

\bibliographystyle{chicago}
\bibliography{frame}
}
\end{document}